\RequirePackage{ifpdf}
\ifpdf
\else
\PassOptionsToPackage{dvipdfmx}{graphicx}
\PassOptionsToPackage{dvipdfmx}{hyperref}
\fi

\documentclass[12pt,a4paper]{article}
\usepackage{jheppub}
\allowdisplaybreaks[4]
\usepackage{amsmath,bm}
\usepackage{subfigure}
\usepackage{slashed}
\usepackage{mathrsfs}
\usepackage{amssymb}                                    		
\usepackage{mathtools}                  		
\usepackage{latexsym}                   		
\usepackage{bbold}
\usepackage{csquotes}
\MakeOuterQuote{"}
\usepackage{comment}

\title{
Chiral symmetry breaking and restoration by helical magnetic fields in AdS/CFT
}

\author[1]{Martí Berenguer,}
\author[1]{Javier Mas,}
\author[2]{Masataka Matsumoto,}
\author[3]{Keiju Murata}
\author[1]{and Alfonso V. Ramallo}
\affiliation[1]{Departamento de Física de Partículas, Universidade de Santiago de Compostela and Instituto Galego de Física de Altas Enerxías (IGFAE). E-15782 Santiago de Compostela, Spain}
\affiliation[2]{Wilczek Quantum Center, School of Physics and Astronomy,
Shanghai Jiao Tong University, Shanghai 200240, China}
\affiliation[3]{Department of Physics, College of Humanities and Sciences, Nihon University, Sakura-josui,
Tokyo 156-8550, Japan}
\emailAdd{marti.berenguer.mimo@usc.es}
\emailAdd{javier.mas@usc.es}
\emailAdd{masataka@sjtu.edu.cn}
\emailAdd{murata.keiju@nihon-u.ac.jp}
\emailAdd{alfonso.ramallo@usc.es}

\abstract{%
We study the effects of helical magnetic fields on chiral symmetry breaking within the AdS/QCD framework using the D3/D7-brane model. By analyzing the brane embeddings, we obtain three types of massless solutions, corresponding to three phases with different behavior in the dual field theory. From the study of quark condensates, free energy, and electric currents, we find that helical magnetic fields can counteract uniform-field-induced symmetry breaking, driving the system towards symmetry restoration. We also find an effect analog to the chiral magnetic effect whereby the current is parallel to the magnetic field. We further study the massive case, and find that the helical configuration is less effective in erasing the first order phase transition that is present in the case of a constant magnetic field.

}

\begin{document}
\maketitle

\section{Introduction}

The AdS/CFT correspondence offers a powerful framework for exploring strongly coupled quantum field theories, including aspects of Quantum Chromodynamics (QCD). This duality between gravitational theories in higher-dimensional spacetimes and gauge theories has been successfully applied for studying non-perturbative phenomena. Among the various holographic models developed, the D3/D7-brane intersection \cite{Karch:2002sh} has emerged as a prominent tool for studying fundamental aspects of QCD, a framework commonly referred to as AdS/QCD. In this model, a probe D7-brane is embedded into an asymptotically AdS spacetime, providing a powerful framework for non-perturbative analyses of gauge theory dynamics, particularly in the presence of external fields and finite temperature.

A key feature of low-energy QCD is the spontaneous breaking of chiral symmetry \cite{Bergman:2012na}. Within the AdS/QCD framework, this phenomenon can be robustly modeled. For instance, it is well known that a uniform magnetic field induces chiral symmetry breaking \cite{Filev:2007gb}.

However, heavy ion collision experiments suggest that the magnetic fields generated in such environments are neither static nor uniform. In particular, helical magnetic fields have been identified as a relevant configuration \cite{Skokov:2009qp,Voronyuk:2011jd,Bzdak:2011yy,Deng:2012pc}. The interplay between such non-uniform magnetic fields and the dynamics of chiral symmetry breaking and restoration remains poorly understood. Therefore, developing methods to incorporate such spatially varying magnetic fields within AdS/QCD is crucial for gaining deeper insight into real-world QCD dynamics.

In this work, we investigate the effects of helical magnetic fields on chiral symmetry dynamics within the AdS/QCD framework. The theory of interest is $\mathcal{N}=4$ SYM in $(3+1)$-dimensions with gauge group $SU(N_c)$, coupled to a number $N_f\ll N_c$ of $\mathcal{N}=2$ flavor hypermultiplets in the fundamental representation of $SU(N_c)$. Holographically, this system is modeled as $N_f$ probe D7-branes embedded in the AdS$_5\times S^5$ geometry sourced by a stack of $N_c$ coincident D3-branes \cite{Karch:2002sh}. Additionally, we include an axial $U(1)_A$ field $A_j^5=b/2~\delta_{jz}$, which describes a Weyl semimetal. The holographic realization of this setup was developed in \cite{Fadafan_2021}. 

Employing the model described above, we introduce a spatially varying magnetic field with nonzero helicity to explore its influence on chiral symmetry breaking and restoration. Our results show that the helical structure of the magnetic field counteracts the symmetry-breaking effects of a constant magnetic field, ultimately driving the system towards chiral symmetry restoration. We are also interested in the possibility that a helical magnetic field could induce the helical structure of the brane, determined by the parameter $b$. We find that this effect is not induced by the helical magnetic field.

The paper is organized as follows: In Section \ref{sec:Setup} we introduce the model and relate the bulk fields to the gauge theory quantities at the boundary. We also outline the different types of solutions encountered. Section \ref{sec:breakingrestoration} focuses on the massless solutions, classifying them according to their chiral symmetry breaking pattern. We compare their free energies, chiral condensates and response currents, demonstrating that sufficiently helical magnetic fields restore chiral symmetry. We extend the analysis to the massive case in Section \ref{sec:massive}. We conclude in Section \ref{sec:conclusions} with an outlook and future directions. Additional details of the derivations and analysis are relegated to Appendices \ref{app:HolRnom} and \ref{app:discretescale}

\section{Introducing a helical magnetic field in the D3/D7 model}\label{sec:Setup}

\subsection{Setup}

The near-horizon geometry sourced by $N_c$ D3-branes is given by AdS$_5 \times S^5$, 
\begin{equation}
 ds^2=\frac{r^2+R^2}{L^2}[-dt^2+dx^2+dy^2+dz^2]
  +\frac{L^2}{r^2+R^2}[dr^2 + r^2 d\Omega_3^2 + dR^2 +R^2d\phi^2]\ ,
  \label{AdS5S5}
\end{equation}
where $L$ is the AdS radius, which we set to unity hereafter. We introduce $N_f\ll N_c$ probe D7-branes in this background. The coordinates $(t,x,y,z)$ are common to both sets of branes, while the remaining 6 transverse directions split into two groups: four coordinates parallel to the D7-branes but orthogonal to the D3-branes, and two transverse to both sets of branes. In \eqref{AdS5S5}, these are written in spherical coordinates and are denoted as $(r,\Omega_3)$ and $(R,\phi)$, respectively.

The dynamics of $N_f$ coincident D7-branes is governed by the Dirac-Born-Infeld (DBI) action,
\begin{equation}
 S=-N_f T_{D7} \int d^8\xi \sqrt{-\textrm{det}[h_{ab}+2\pi\alpha' F_{ab}]}\ ,
\label{SDBI}
\end{equation}
where $T_{D7}$ is the D7-brane tension $h_{ab}$ is the induced metric, and $F_{ab}=\partial_a A_b-\partial_b A_a$ is the $U(1)$-gauge field strength. The Wess-Zumino term vanishes in the setup under consideration. The worldvolume coordinates are $\xi^a=(t,x,y,z,r,\Omega_3)$. We assume spherical symmetry in $S^3$ as well as translational symmetry in the $(t,x,y)$ directions, while we allow for a breaking of symmetry along the $z$-direction.

To introduce a helical magnetic field, we consider the following ansatz\footnote{
In Refs.~\cite{Hashimoto:2016ize,Kinoshita:2017uch,Ishii:2018ucz,Garbayo:2020dmh,Berenguer:2022act}, 
a rotating electric field in the D3/D7 and D3/D5 systems was studied with the ansatz $A_x + iA_y \propto e^{i\Omega t}$. With the help of this idea, we made the ansatz in Eq.~(\ref{aRdef}). We will see that the equations of motion of the brane are given by ordinary differential equations even though the gauge field is $z$-dependent.
}
\begin{equation}
    2\pi \alpha' (A_x + iA_y) = a(r) e^{ikz}\ , \quad R = R(r)\ , \quad \phi = bz\ .
\label{aRdef}
\end{equation}
This choice explicitly breaks translational symmetry along the $z$-direction. Although we could include an $r$-dependence for $\phi$ as $\phi = bz + \Phi(r)$, any such dependence leads to a nonzero imaginary part in the action, signalling a tachyonic instability \cite{Fadafan_2021}, which ultimately forces $\Phi(r)$ to be constant.
Substituting the above ansatz into the DBI action~(\ref{SDBI}), we obtain
\begin{align}
    &S= \mathcal{N} V_4 \int dr \mathcal{L}_0\ ,\\
    &\mathcal{L}_0\equiv-\frac{r^3}{r^2+R^2}
\left[
(1+a'{}^2 + R'{}^2)(k^2 a^2 + b^2 R^2 + (R^2+r^2)^2)
\right]^{1/2}\ ,
\end{align}
where $\mathcal{N}\equiv 2\pi^2 N_f T_{D7}$ and $V_4=\int dtdxdydz$. 
The equations of motion derived from this action are
\begin{align}
    &a''=-\frac{1+a'{}^2+R'{}^2}{r (r^2+R^2) (k^2 a^2 + b^2 R^2 + (r^2+R^2)^2)}\notag\\
&\qquad \times\big[
 a'(r^2+3R^2) (k^2 a^2 + b^2 R^2)
+ 3 a'( r^2+R^2)^3 
- k^2r a (r^2 + R^2) 
\big]\ ,\label{beq}\\
    &R''=-\frac{1+a'{}^2+R'{}^2}{r (r^2+R^2) (k^2 a^2 + b^2 R^2 + (r^2+R^2)^2)}\notag\\
&\qquad \times\big[
R'(r^2+3 R^2)  (k^2 a^2 + b^2  R^2 )
+ 2 k^2 r a^2 R \notag\\
&\qquad\qquad\qquad\qquad
+ b^2 r R(R^2 - r^2) 
+ 3 R' (r^2+R^2)^3 
)
\big]
\ .\label{weq}
\end{align}
We solve these equations numerically and determine the brane profile.

\subsection{Physical quantities in the boundary theory}

Near the AdS boundary, $r\to\infty$, the solutions to Eqs. (\ref{beq}) and (\ref{weq}) can be expanded as
\begin{equation}
\begin{split}
    &a(r)=a_\infty + \frac{J}{2r^2}  -\frac{k^2 a_\infty }{2r^2} \log\left(\frac{r}{\sqrt{k a_\infty}}\right) + \cdots\ ,\\
    &R(r)=M  + \frac{C}{r^2}  -\frac{b^2 M}{2r^2} \log\left(\frac{r}{M}\right) + \cdots\ .
    \end{split}
\label{aRexpand}
\end{equation}
Logarithmic terms appear in the asymptotic expansion. In their arguments, we have normalized the radial coordinate $r$ by the constants $\sqrt{k a_\infty}$ and $M$. This normalization can always be done by redefining $J$ and $C$. As we will see shortly, with the above definition of $J$ and $C$, these quantities remain covariant under scaling transformations.

The coefficients in the expansion are related to the physical quantities in the boundary theory. For example, from Eq.~(\ref{aRdef}) and Eq.~(\ref{aRexpand}), the asymptotic value of the gauge field, which corresponds to the external gauge field in the boundary theory, is given by $(A_x + i A_y)|_{r\to\infty} = a_\infty e^{ikz}/(2\pi\alpha')$. The magnetic field in the boundary theory is given by $\bm{B}_\textrm{QFT} = \nabla \times (A_x, A_y, A_z)|_{r\to\infty}$. Its explicit expression is 
\begin{equation}
 B_x + i B_y|_\textrm{QFT} = -\frac{\sqrt{\lambda}}{2\pi} k a_\infty e^{ikz} \ ,\quad B_z|_\textrm{QFT} = 0\ .
 \label{BQFT}
\end{equation}
We express the string scale as $\alpha' = 1/\sqrt{\lambda}$, where $\lambda$ is the 't Hooft coupling. (Recall that we have set the AdS radius to $L = 1$). The gauge field above describes a helical magnetic field: for each $(x,y)$-plane, the magnetic field is homogeneous, but its direction rotates as we vary the coordinate $z$. We define 
\begin{equation}
 B = k a_\infty~,
 \label{eq:defmag}
\end{equation}
which characterizes the amplitude of the helical magnetic field in the boundary theory. Since the DBI action is invariant under $a(r) \to -a(r)$, we can assume $B \geq 0$ without loss of generality. A uniform magnetic field, independent of $z$, is recovered in the double-scaling limit $k \to 0$, while keeping $B$ finite. This limit corresponds to the setup considered in \cite{Evans:2024ilx}.

It was shown in \cite{Fadafan_2021} that the parameter $b$ corresponds to a component of the non-dynamical $U(1)_A$ gauge field, $A_\mu^5|_\textrm{QFT}$. The other parameters, $(M, J, C)$, are related to the quark mass $M_\textrm{QFT}$, the electric current $\bm{J}_\textrm{QFT}$, and the quark condensate $\langle \mathcal{O}_m \rangle|_\textrm{QFT}$ in the boundary theory, respectively. The correspondence is summarized as follows:
\begin{equation}
\begin{aligned}
    M_\textrm{QFT} & = \frac{\sqrt{\lambda}}{2\pi} M~, & \quad A^5_\mu|_\textrm{QFT} & =\frac{1}{2}\partial_\mu\phi=\frac{b}{2}\delta_\mu^z\\
    \langle \mathcal{O}_m \rangle|_\textrm{QFT} & = \frac{N_c N_f \sqrt{\lambda}}{4\pi^3} \left(-C + \frac{b^2 M^2}{4}\right)~, & & \\
    J_x + i J_y|_\textrm{QFT} & = \frac{N_c N_f \sqrt{\lambda}}{(2\pi)^3} \left(J + \frac{k B}{4}\right) e^{ikz}~, & \quad J_z|_\textrm{QFT} & = 0
    \label{PhysicalQ}
\end{aligned}
\end{equation}
A detailed derivation of $\bm{J}_\textrm{QFT}$ and $\langle \mathcal{O}_m \rangle|_\textrm{QFT}$ is provided in Appendix \ref{app:HolRnom}.
In the massless case $M=0$, the parameter $b$ can be removed by a field redefinition in the boundary theory and becomes irrelevant.

We also introduce the free energy as
\begin{equation}
 f = -\frac{S + S_\textrm{ct}}{\mathcal{N} V_4} \ ,
 \label{fdef}
\end{equation}
where $S$ is the on-shell DBI action (\ref{SDBI}) and $S_\textrm{ct}$ contains the counterterms, whose explicit expressions are given in Appendix \ref{app:HolRnom}. The free energy will be used to determine the most stable solution.

The equations of motion are invariant under the following scaling transformations:
\begin{equation}
\begin{aligned}
    (t, x, y, z) & \to \lambda^{-1} (t, x, y, z)~,  & r & \to \lambda r ~, & (k, b) & \to \lambda (k, b)~,\\
    R(r) & \to \lambda R(r)~, & a(r) & \to \lambda a(r)~, & &
\end{aligned}
\end{equation}
where $\lambda$ is an arbitrary positive constant. Additionally, the transformation laws for quantities in the asymptotic expansion are given by
\begin{equation}
\begin{aligned}
    M & \to \lambda M~,  & C & \to \lambda^3 C ~, & a_\infty & \to \lambda a_\infty~,\\
    J & \to \lambda^3 J~, & B & \to \lambda^2 B~, & f & \to \lambda^4 f~.
    \label{scalsym}
\end{aligned}
\end{equation}
We will consider scale-invariant combinations of parameters. We will mostly normalize physical quantities by the magnetic field $B$, since we are interested in the physics of a nonzero helical magnetic field.

\subsection{Typical solutions}\label{sec:Solutions}

In this section, we numerically obtain the solutions for the fields in the brane worldvolume. The gravitational background is pure AdS$_5\times S^5$ and there is no event horizon. 

\subsubsection{Minkowski embeddings}

In this case there  is also no singular shell~\cite{Karch:2007pd}, which corresponds to the event horizon of the effective metric of the brane: $\bar{h}_{ab}=h_{ab}+(2\pi\alpha')^2 h^{cd} F_{ac}F_{bd}$~\cite{Seiberg:1999vs,Gibbons:2000xe,Gibbons:2001ck,Gibbons:2002tv,Kim:2011qh,Hashimoto:2014yza,Hashimoto:2016ize,Kinoshita:2017uch}. In fact, the denominators of right hand sides of Eqs.~(\ref{beq}) and (\ref{weq}) are always nonzero, meaning that all solutions extend to the axis $r\to 0$. 
Here, we focus on the case where $a_0\equiv a(0)\neq 0$ and $R_0\equiv R(0)\neq 0$. 
For $r\sim 0$, the solutions behave as
\begin{equation}
\begin{split}
    &a(r)=a_0+\frac{k^2 a_0}{8(k^2 a_0^2+ b^2 R_0^2 + R_0^4)}r^2+\cdots, \\
    &R(r)=R_0-\frac{(2 k^2 a_0^2 + b^2 R_0^2)}{8 R_0 (k^2 a_0^2 + b^2 R_0^2 + R_0^4)}r^2+\cdots \ .
\end{split}
\label{Mink_axis}
\end{equation}
We integrate Eqs.~(\ref{beq}) and (\ref{weq}) numerically from $r=r_\textrm{min}$ to $r=r_\textrm{max}$, choosing $r_\textrm{min}\sim 10^{-4}$ and $r_\textrm{max}\sim 10^3$ in our numerical calculations.

\begin{figure}[t]
   \centering
  {\includegraphics[width=0.45\textwidth]{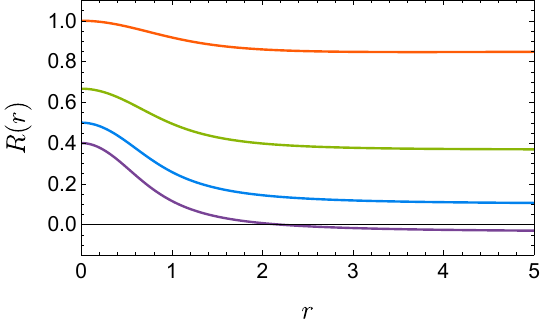}\label{Rtypical}
   }
  {\includegraphics[width=0.45\textwidth]{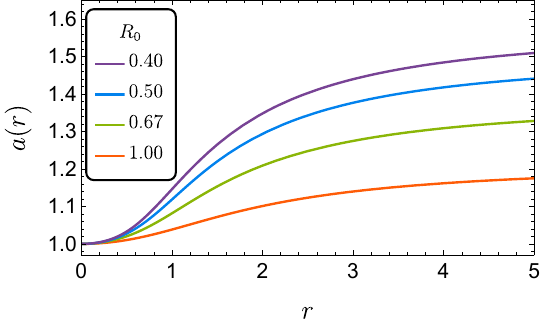}\label{atypical}
   }
  \caption{Typical solutions of $R(r)$ (left) and $a(r)$ (right) for the Minkowski embedding. We show the solutions for several values of $R_{0}$ while the other parameters are fixed as $k=b=a_{0}=1$.}
  \label{fig:Minsols}
 \end{figure}

Fig. \ref{fig:Minsols} shows typical solutions under the above boundary conditions. Here, we vary $R_0$ as $R_0= \{0.40, 0.50, 0.67, 1.00\}$ while the other parameters are fixed as $k=b=a_0=1$. Following the standard nomenclature, we refer to this type of solutions as "Minkowski embeddings".

As $R_0$ is lowered, the brane crosses the horizontal axis at a specific value of $R_0$, as shown in the left plot of Fig.~\ref{fig:Minsols}. According to the asymptotic behaviour \eqref{aRexpand}, the quark mass becomes zero for the value of $R_{0}$ at which the brane crosses the axis asymptotically at $r\to\infty$.

Since the asymptotic value of $a(r)$ at the boundary, namely $a_{\infty}$, gives the amplitude of the helical magnetic field via Eq.~(\ref{eq:defmag}), we observe from the right plot in Fig.~\ref{fig:Minsols} that decreasing $R_{0}$ corresponds to increasing $B$. This crossing behaviour has already been found in the case that a homogeneous magnetic field is applied~\cite{Bergman:2012na}. 

\subsubsection{Exponential embeddings}
\begin{figure}[t]
   \centering
    \includegraphics[width=0.45\textwidth]{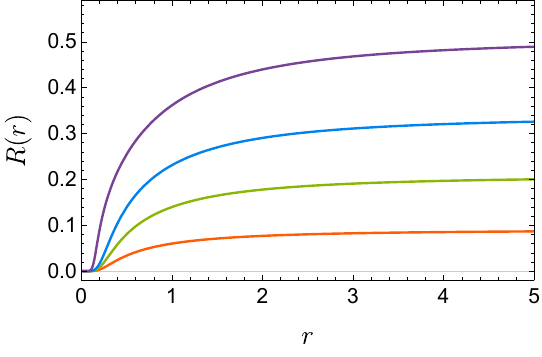}
    \includegraphics[width=0.45\textwidth]{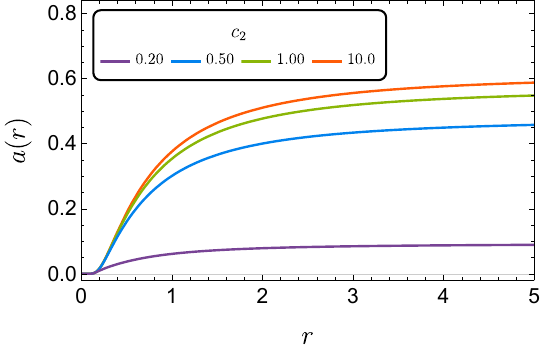}
  \caption{Typical solutions of $R(r)$ (left) and $a(r)$ (right) for the exponential embedding. We show the solutions for several values of $c_{2}$ while the other parameters are fixed as $k=b=c_{1}=1$. }
  \label{fig:Expsols}
 \end{figure}

For $a_0=R_0=0$, we easily find the trivial solution $a(r)=R(r)=0$. However, as shown in Ref.~\cite{Fadafan_2021}, there exists a family of non-trivial solutions whfor which both $a(r)$ and $R(r)$ approach $0$ as $r\to 0$. To find these solutions, we assume that $a(r)$ and $R(r)$ are sufficiently small near the axis $r=0$ and we consider the linearized equations of motion. Substituting $a(r)\to \epsilon a(r)$ and $R(r)\to \epsilon R(r)$ into Eqs.~(\ref{beq}) and (\ref{weq}), and expanding to first order in $\epsilon$, we obtain the following decoupled equations
\begin{equation}
    a''+\frac{3}{r}a'-\frac{k^2}{r^4}a=0\ ,\quad
    R''+\frac{3}{r}R'-\frac{b^2}{r^4}R=0\ .
\end{equation}
The regular solutions at $r=0$ are given by $a\propto K_1(k/r)/r$ and $R\propto K_1(b/r)/r$, where $K_\alpha(x)$ is the modified Bessel function of the first kind. In the vicinity of $r = 0$, these solutions can be approximated as
\begin{equation}
    a(r)\simeq  \frac{c_1}{\sqrt{r}}e^{-k/r}\ ,\quad 
    R(r)\simeq \frac{c_2}{\sqrt{r}}e^{-b/r}\ .\label{expem}
\end{equation}
Solutions satisfying these boundary conditions are referred to as "exponential embeddings". Typical profiles of these solutions are shown in Fig.~\ref{fig:Expsols}. 
Note that for the exponential embedding the solutions with $M=0$ are possible only when $R$ is the trivial embedding with $c_2=0$, which means there are no chiral symmetry breaking solutions of this type.

\begin{figure}[h]
   \centering
    \includegraphics[width=0.32\textwidth]{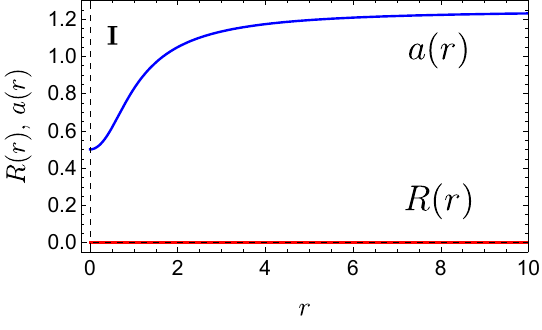}
    \includegraphics[width=0.32\textwidth]{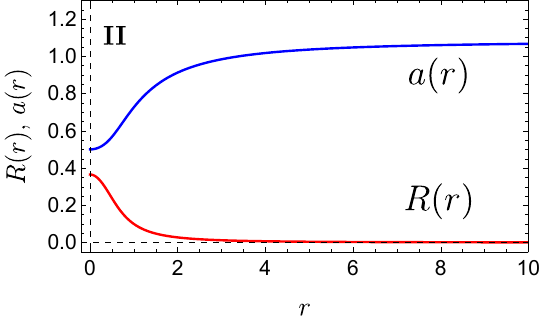}
    \includegraphics[width=0.32\textwidth]{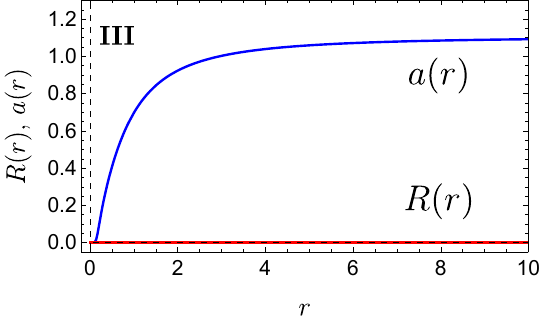}
  \caption{Typical solutions of $a(r)$ and $R(r)$ for type I (left), II (middle), and III (right) with $b=0$.}
  \label{fig:3type}
 \end{figure}

\section{Chiral symmetry breaking and its restoration}\label{sec:breakingrestoration}

A well-known effect of introducing a worldvolume magnetic field is the possibility of triggering spontaneous chiral symmetry breaking \cite{Babington_2003,Filev:2007gb}. We observe a similar behaviour here; however, we must now consider the two types of solutions we have found so far. 

\subsection{Classifying solutions for \texorpdfstring{$M=0$}{M=0}}

In this section, we only focus on the massless case, $M=0$, and study spontaneous chiral symmetry breaking in the presence of a helical magnetic field. We classify the solutions into three types:
\begin{description}
\item[I] Chiral symmetry preserving Minkowski embeddings: 
\[
 a(r)\in F_\textrm{Mink}\ ,\quad  R(r)\equiv 0\ .
\]
\item[II] Chiral symmetry breaking Minkowski embeddings:
\[
a(r)\in F_\textrm{Mink}\ ,\quad  R(r)\in F_\textrm{Mink}\ .
\]
\item[III] Chiral symmetry preserving exponential embeddings: 
\[
 a(r)\in F_\textrm{Exp}\ ,\quad  R(r)\equiv 0\ .
\]
\end{description}
Here, $F_\textrm{Mink}$ and $F_\textrm{Exp}$ are sets of non-trivial functions which behave as described by Eq.~(\ref{Mink_axis}) and (\ref{expem}) near the axis, respectively. 
Typical profiles of types I, II and III are shown in Fig.~\ref{fig:3type}.  

\begin{figure}[h!]
   \centering
    \includegraphics[width=0.49\textwidth]{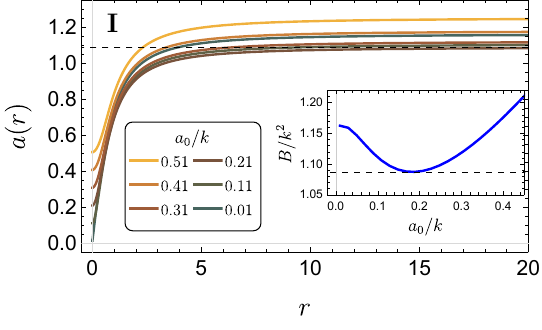}
    \includegraphics[width=0.49\textwidth]{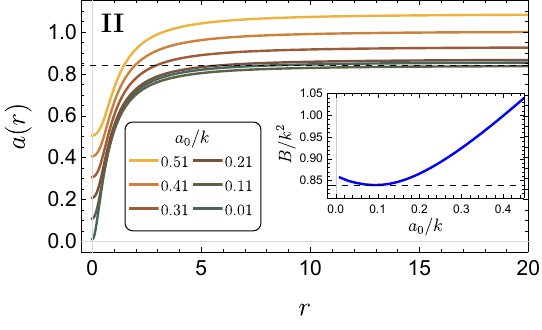}\\
    \includegraphics[width=0.49\textwidth]{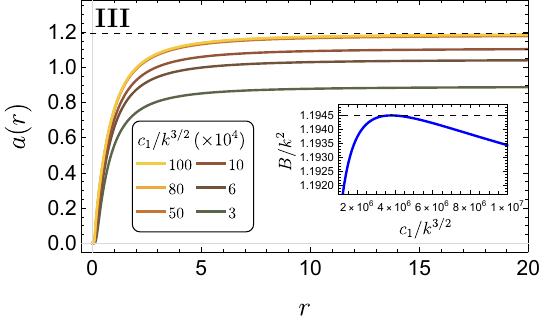}
  \caption{Typical solutions of $a(r)$ for type I (left), II (middle), and III (right) with $b=0$. The inset shows the value of the magnetic field as a function of the parameters, $a_0$ for type I and II, and $c_1$ for type III. The dashed line denotes the lower bound of $B/k^{2}$ for type I and II, and the upper bound for type III.}
  \label{fig:3type2}
\end{figure}

For type II solutions, the brane profile $R(r)$ is non-trivial but asymptotically approaches zero as $r\to\infty$. In this case, the quark condensate is non-zero and chiral symmetry is spontaneously broken. 

In contrast, for type III solutions, where $a(r)\in F_\textrm{Exp}$ and $R(r)\in F_\textrm{Exp}$, there is no massless solution. Therefore, for type III, the brane profile $R(r)$ must correspond to the trivial solution. 

Fig.~\ref{fig:3type2} shows the profiles of $a(r)$ for type I (left), II (middle), and III (right), with the parameters normalized by $k$. In each panel, we vary the value of the parameter $a_{0}/k$ for types I and II, and $c_{1}$ for type III. The inset shows the magnetic field $B/k^{2}$ as a function of these parameters.

As indicated by the black dashed line, we observe that there is a lower bound of $B/k^{2}\gtrsim 1.09$ for type I, and of $B/k^{2}\gtrsim 0.84$ for type II, and an upper bound $B/k^{2}\lesssim 1.1945$ for type III. In other words, for a fixed magnetic field $B$, the wavenumber $k$ is bounded by $k/\sqrt{B}\lesssim 0.958$ and $1.091$ for type I and II, respectively, and $k/\sqrt{B}\gtrsim0.915$ for type III. These bounds for each type of solutions can be visually confirmed by the plots of free energy, quark condensate, and electric current in the following section.

\begin{figure}[h!]
\begin{center}
\includegraphics[width=0.7\textwidth]{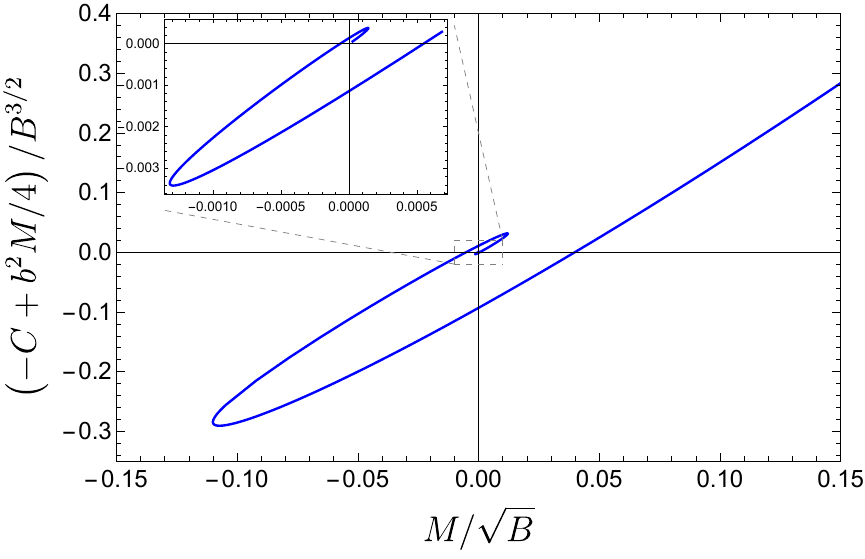}
\end{center}
\caption{The quark condensate $-C+b^{2}M/4$ is plotted as a function of the quark mass $M$, normalized by the magnetic field $B$. 
The value of $R_{0}$ decreases from the outer to the inner regions of the spiral, with $k=b=a_{0}=1$ fixed. The inset provides a zoom-in on the region near the origin.}
\label{fig:c_vs_m}
\end{figure}

Fig.~\ref{fig:c_vs_m} shows the quark condensate as a function of the quark mass for type II solutions, normalized by the magnetic field $B$.
The value of $R_{0}$ decreases from the outer to the inner regions of the spiral in the plot, with the other parameters fixed to $k=b=a_{0}=1$. The inset of Fig.~\ref{fig:c_vs_m} reveals a self-similar spiral behaviour near the origin, indicating the instability of the embedding around $R_{0}=0$. Note that only the probe brane profile approaches $R=0$ as $R_{0}\to 0$; the profile of $a(r)$ is still of Minkowski type, with $a_{0}=1$ fixed. A similar spiral behaviour has been observed in the case of a homogeneous magnetic field \cite{Filev:2007gb}.

This spiral behavior implies the existence of an infinite number of non-trivial solutions, even in the massless case. These solutions generally have non-zero quark condensate and correspond to chiral symmetry-breaking phases in the dual field theory. A more detailed analysis of this spiral behavior can be found in Appendix \ref{app:discretescale}.

\begin{figure}[h!]
   \centering
    \includegraphics[width=0.6\textwidth]{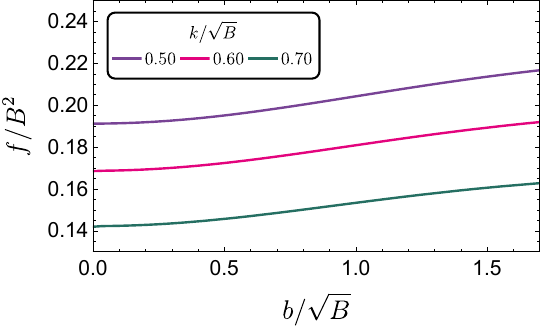}
  \caption{Free energy $f/B^2$ as a function of $b/\sqrt{B}$ for type II solutions. We show three different values for the wavenumber of the helical magnetic field.}
  \label{f_vs_k_b}
\end{figure}

\subsection{Comparing free energies}
For the massless case $M=0$, solutions are parameterized by two wavenumbers $k$ and $b$, with $B$ used for normalization. Note that in the massless case, the value of $b$ is not determined by the boundary condition at infinity, \textit{i.e.}, it cannot be considered a source in the boundary theory. By introducing cartesian coordinates $(x_8, x_9) = (R \cos \phi, R \sin \phi)$, the boundary condition for the brane can be written as $(x_8, x_9)|_{r \to \infty} = (M \cos(bz), M \sin(bz))$.

For $M \neq 0$, we can control the parameter $b$ by imposing the appropriate boundary condition at infinity. However, when $M = 0$, the boundary condition becomes trivial, and we can no longer use it to control the value of $b$\footnote{The parameter $b$ can be removed by a chiral rotation of the Dirac field in the boundary theory when $M=0$~\cite{Fadafan_2021}}. Consequently, $b$ will be determined dynamically. We fix $k/\sqrt{B}$ and search for the value of $b/\sqrt{B}$ that minimizes the free energy.

Here, we compute the free energy for type II solutions as a function of these two parameters. As mentioned earlier, there are infinitely many type II solutions that share the same wavenumbers $(k/\sqrt{B}, b/\sqrt{B})$. Here, we focus only on the zero-node solution, as it is found to have the minimum free energy among type II solutions. For type I and type III solutions, the brane profile $R(r)$ is trivial, and their free energies are independent of $b$. We will consider their free energies shortly. Fig. \ref{f_vs_k_b} shows the free energy $f/B^2$ for type II solutions as a function of $b/\sqrt{B}$ for $k/\sqrt{B}=0.5, 0.6$ and $0.7$. For a fixed value of $k/\sqrt{B}$, the brane achieves minimum free energy when $b/\sqrt{B}=0$. This indicates that solutions with $b=0$ are "thermodynamically" favored. In other words, even in the presence of a helical magnetic field, the helical structure of the brane embedding is not induced.
\begin{figure}[h!]
   \centering
    \includegraphics[width=0.6\textwidth]{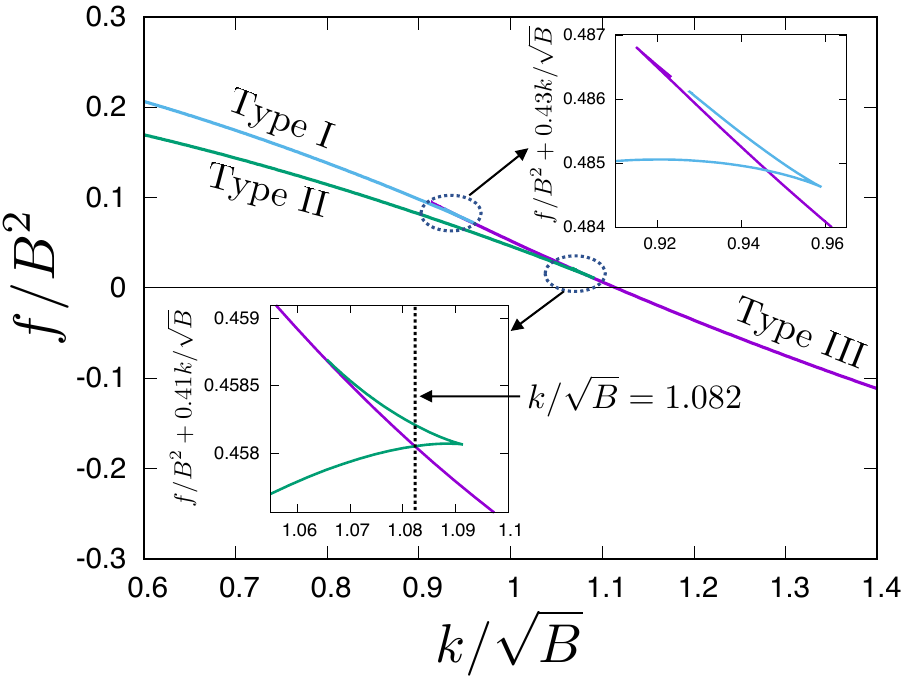}
  \caption{Free energy $f/B^2$ as a function of $k/\sqrt{B}$ for type I, II and III solutions. For type II solutions, we set $b/\sqrt{B}=0$. The inset shows a close-up view of the region around the connection points between type I and type III, as well as between type II and type III. To enhance visibility, the vertical axes in the insets are adjusted to $f/B^2 + \alpha k/\sqrt{B}$ with $\alpha = 0.41$ and $0.43$.} 
  \label{f_vs_k}
\end{figure}

We now examine the free energy of solutions with $b=0$. Fig. \ref{f_vs_k} shows the free energy $f/B^2$ as a function of $k/\sqrt{B}$ for type I, II, and III solutions. The insets show magnified views of the vicinity of the connection points between type I and type III, as well as between type II and type III. To enhance visibility, the vertical axes in the insets are rotated to $f/B^2 + 0.41k/\sqrt{B}$ and $f/B^2 + 0.43k/\sqrt{B}$, respectively. All curves for the free energy exhibit a folding behavior at the connection points.  

Type I solutions always have a larger free energy than type II solutions, and thus they cannot be the favored solutions.
As observed earlier, type II and type III solutions exist for $k/\sqrt{B} \lesssim 1.09$ and $k/\sqrt{B} \gtrsim 0.915$, respectively. Type II and III solutions have the smallest free energy for $k/\sqrt{B}<1.082$ and $k/\sqrt{B}>1.082$, respectively, indicating a phase transition between type II and type III at $k/\sqrt{B}=1.082$.

\begin{figure}[h!]
   \centering
    \includegraphics[width=0.6\textwidth]{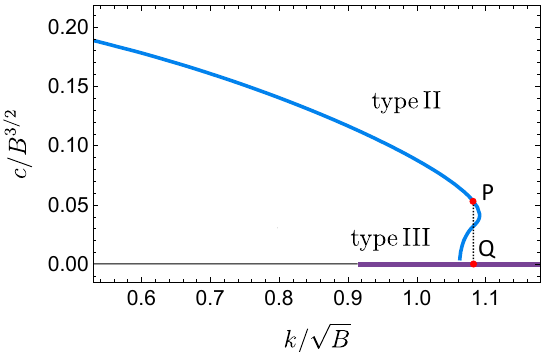}
  \caption{The quark condensate for the type II and III solutions. Points P and Q are the transition points. There is a discontinuous change of the quark condensate due to the transition.}
  \label{fig:kcplot}
\end{figure}

Fig.~\ref{fig:kcplot} shows the quark condensate for type II and type III solutions. Points P and Q mark the transition points. The phase transition causes a discontinuous change in the quark condensate. Since type III solutions have zero quark condensate, the restoration of chiral symmetry for $k/\sqrt{B}>1.082$ is evident in this figure.

\subsection{On the helical electric current}

From Eqs.(\ref{BQFT}) and (\ref{PhysicalQ}), we obtain an electric current proportional to the helical magnetic field:
\begin{equation}
 \bm{J}|_\textrm{QFT}=-\frac{N_c N_f}{(2\pi)^2}
\left(\frac{J}{B} + \frac{k}{4}\right) \bm{B}|_\textrm{QFT}\ .
\end{equation}

 \begin{figure}[h!]
   \centering
    \includegraphics[width=0.5\textwidth]{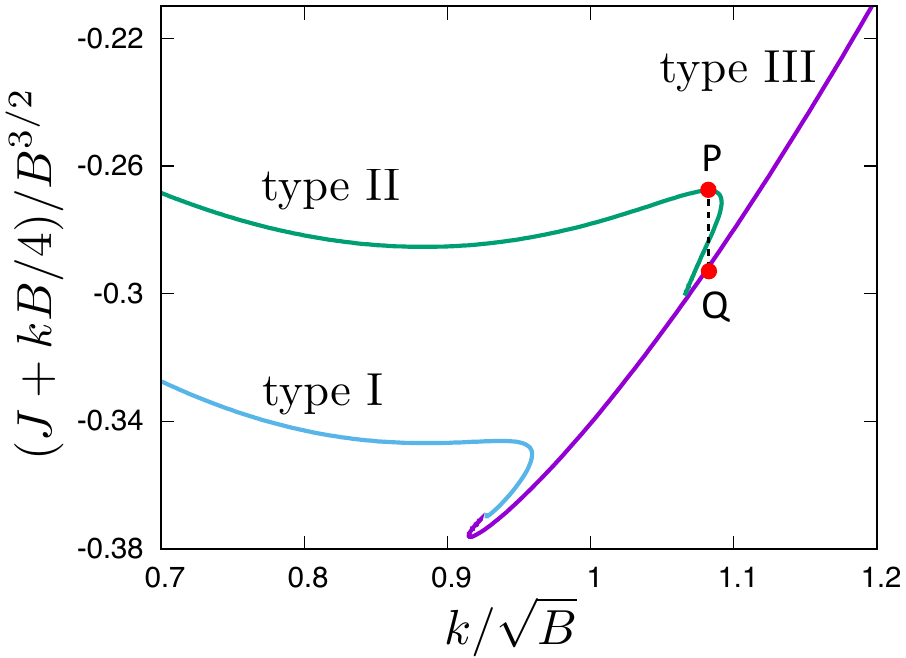}
  \caption{The electric current $(J+kB/4)/B^{3/2}$ as a function of the wavenumber $k/\sqrt{B}$.
  Points P and Q mark the transition points. The phase transition induces a discontinuous change in the current. 
   }
  \label{fig:kJplot}
\end{figure}

This is a somewhat surprising prediction of the holographic model: there is a non vanishing electric current parallel to the magnetic field in each plane. Fig.~\ref{fig:kJplot} shows the electric current $(J+k B/4)/B^{3/2}$ as a function of the wavenumber $k/\sqrt{B}$. Again, we focus on the massless case: $M=0$. The dashed vertical line represents the critical wavenumber of the phase transition: $k/\sqrt{B}=1.082$. Points P and Q are the transition points, where the current undergoes a discontinous change, similar to the quark condensate.

The microscopic origin of this electric current remains an open question, since we are here in the confined phase and mesons are expected to be electrically neutral. A more detailed analysis of this effect will be adressed in the future.

\section{Massive case}\label{sec:massive}

In the absence of a magnetic field, the model exhibits a first-order phase transition from a Weyl semimetal to a trivial insulator as the ratio $|m/b|$ increases \cite{Fadafan_2021}. When a constant magnetic field $B_x$ is applied in the $x$-direction \cite{Evans:2024ilx}, the first order phase transition persists for small values of $B_x$. However, as $B_x$ increases, the transition region gradually shrinks and eventually vanishes, indicating the existence of a critical point along the line of first order transitions. In this section we explore the impact of a helical magnetic field on this scenario. In the limit $k\rightarrow 0$, with $k a_{\infty}=B$ fixed, our results reduce to those of \cite{Evans:2024ilx}.

To facilitate comparison with \cite{Evans:2024ilx}, we normalize all quantities by $b$, setting $b=1$ throughout this section. In these units, the critical point occurs at $B_x=B_c\simeq 0.06$.

\begin{figure}[h!]
    \centering
    \includegraphics[width=0.46\textwidth]{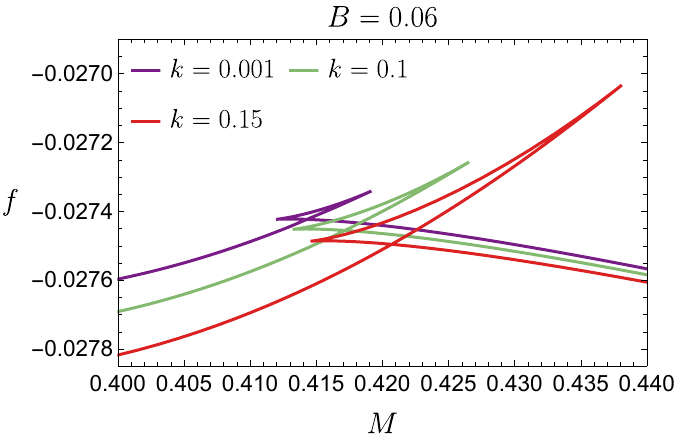}
    \includegraphics[width=0.51\textwidth]{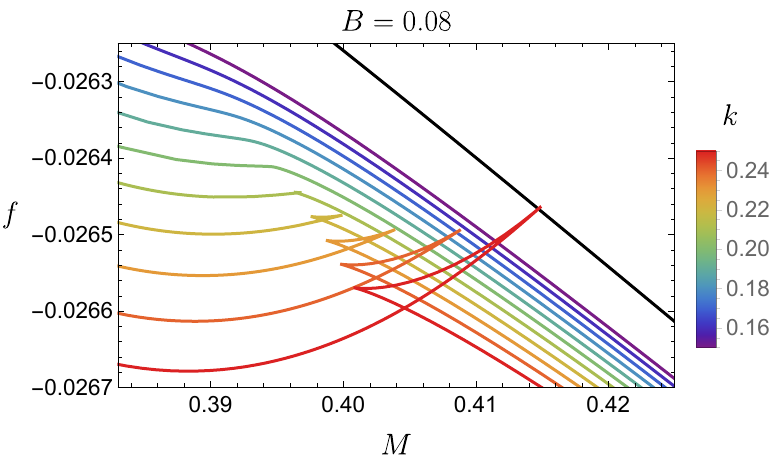}
    \caption{Free energy for two values of the magnetic field. On the left, for $B\lesssim B_c$, we observe that as the phase $k$ increases, the swallow tail structure becomes more pronounced. On the right, for $B=0.08>B_c$, the swallow tail structure emerges when the magnetic field is sufficiently helical, around $k\sim 0.21$. The black line corresponds to the constant $B$ field limit of \cite{Evans:2024ilx}.}
    \label{fig:fmassive}
\end{figure}

Our findings show that, above $B_c$, the helical structure of the magnetic field can reintroduce the phase transition. This behavior is illustrated in Fig.~\ref{fig:fmassive}.

\begin{figure}[h!]
    \centering
    \includegraphics[width=0.49\textwidth]{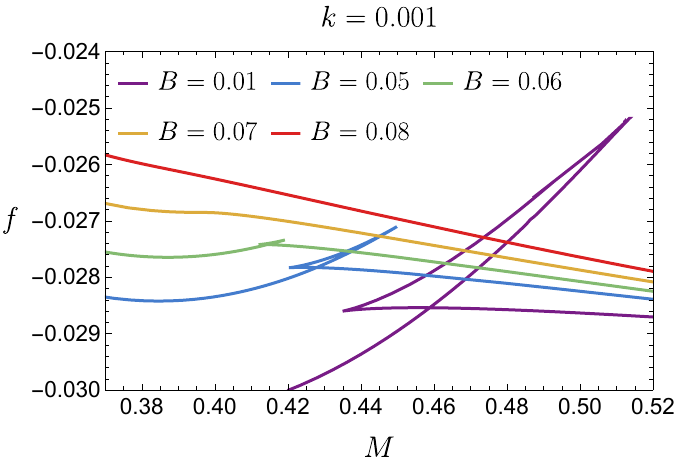}
    \includegraphics[width=0.49\textwidth]{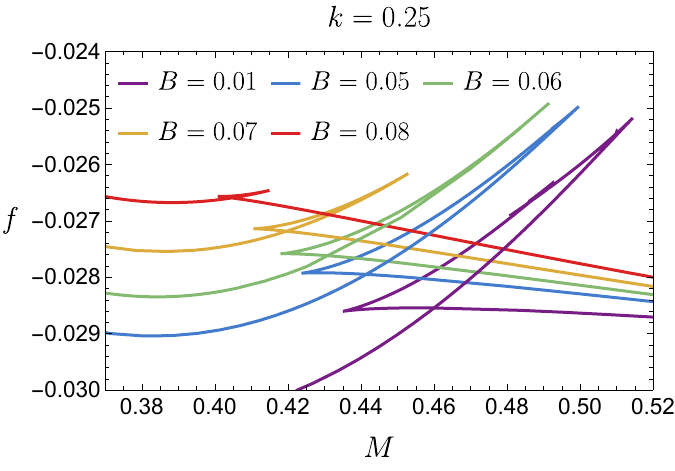}
    \caption{Alternatively, we show the free energy for two values of $k$. On the left, we show the limit of a constant magnetic field from \cite{Evans:2024ilx}. Around $B_c\simeq 0.06$ the phase transition disappears. On the right, we show the free energy for the same values of $B$, but now with a helical structure ($k=0.25$). We observe that the swallow-tail structure of the first order phase transition is enhanced, even for $B>B_c$.}
    \label{fig:fmassivek}
\end{figure}

\section{Conclusions}\label{sec:conclusions}

In this paper we have explored the effects of introducing a helical magnetic field in a D3/D7-brane model holographically dual to a Weyl semimetal. We first examined the case of massless flavor branes, where constant magnetic fields are known to induce chiral symmetry breaking. Our results show that helical magnetic fields counteract the symmetry-breaking effects and drive the system towards chiral symmetry restoration. Initially, we expected the brane embedding to undergo an unstable transition to a helicoidal structure, similar to that of the magnetic field itself. However, our thermodynamic analysis reveals the opposite: the stable configuration is that of a non-helical brane. Thus, from the perspective of the brane embeddings, there are only two such solutions: symmetry-preserving and symmetry-breaking solutions, similar to the case of constant magnetic fields. Notably, the gauge potential admits a class of new solutions, akin to those found in \cite{Fadafan_2021}, which we also term "exponential embeddings" with a little abuse of language. Together, these solution combine to form three distinct phases, with first order phase transitions between them, characterized by the familiar self-similar spiraling behavior seen in many flavor-brane systems. The helical magnetic field also induces a parallel electric current on the flavor degrees of freedom, whose microscopic origin is not fully understood.

In the final part of this paper, we have examined  the massive case. Here, the helical structure of the brane embeddings can be enforced as a boundary condition rather than being dynamically chosen. Our work extends the findings of \cite{Evans:2024ilx}, where a constant magnetic field was shown to erase the first order phase transition between exponential embeddings and Minkowski embeddings.

Future investigations could extend this analysis to finite-temperature effects, which are crucial for understanding environments such as the early-universe or heavy-ion collisions. Time-dependent helical magnetic fields also present an exciting avenue forstudy, as they could model dynamically evolving systems and offer insights into transient phenomena in non-equilibrium settings like heavy ion collisions.

Generalizing the findings to other holographic models or incorporating more realistic QCD-like dynamics would further establish the universality of these phenomena. Exploring the coupling of helical magnetic fields with additional degrees of freedom, such as scalar or pseudoscalar fields, could uncover deeper connections between field configurations and symmetry-breaking patterns.

In summary, our study highlights the rich phenomenology associated with spatially tunable external fields in holographic models, demonstrating their relevance in exploring the phase space of strongly coupled quantum field theories.


\section*{Acknowledgements}
We would like to thank warmly Nick Evans, Takaaki Ishii and Shunichiro Kinoshita and for useful discussions.

This work has received financial support from the Xunta de Galicia (CIGUS Network of Research Centres and grant ED431C-2021/14), the European Union, the María de Maeztu grant CEX2023-001318-M funded by MICIU/AEI/10.13039/501100011033 and the Spanish Research State Agency (grants PID2020-114157GB-I00 and PID2023-152148NB-I00).

The work of MB has been funded by Xunta de Galicia through the Programa de axudas \'a etapa predoutoral da Xunta de Galicia (Conseller\'ia de Cultura, Educaci\'on e Universidade) and the grant 2023-PG083 with reference code ED431F 2023/19.
M.\ M.\ is supported by Shanghai Post-doctoral Excellence Program (No.\,2023338).
The work of K.\ M.\ was supported in part by JSPS KAKENHI Grant Nos.\ JP20K03976, JP21H05186 and JP22H01217.

\appendix

\section{Holographic renormalization}\label{app:HolRnom}

\subsection{On-shell action and counter terms}

In this section, we write the metric of AdS$_5\times S^5$ as
\begin{equation}
    ds^2=\frac{-dt^2+du^2 + dx^2+ dy^2+ dz^2}{u^2}
    +d\theta^2 + \sin^2\theta d\phi^2 + \cos^2\theta d\Omega_3^2\ ,
\end{equation}
where we have set $L=1$. The relation to the coordinates in Eq.~(\ref{AdS5S5}) is given by
\begin{equation}
    R=\frac{1}{u}\sin\theta\ ,\quad 
    r=\frac{1}{u}\cos \theta\ .
\end{equation}

We consider the brane embedding to be of the form $\theta=\theta(u)$ and $\phi=bz$. For the gauge field, we have the ansatz $(2\pi\alpha') (A_x+iA_y) = a(u)e^{ikz}$. In these coordinates, the DBI action is given by
\begin{equation}
\begin{split}
&S=\mathcal{N}V_4 \int du \mathcal{L}\ ,\\
    &\mathcal{L}\equiv\frac{\cos^3\theta}{u^5}\sqrt{(1+u^4a'{}^2+u^2\theta'{}^2)(1+k^2 u^4 a^2+b^2u^2\sin^2\theta)}~.
\end{split}
\label{SDBI_u}
\end{equation}
In this appendix, primes denote derivatives with respect to $u$. The asymptotic expansion of these variables are given by
\begin{equation}
    \begin{split}
        &\theta=Mu + \left(C+\frac{M^3}{6}+\frac{b^2M}{2}\log(Mu)\right)u^3 + \cdots\ ,\\
    &a=a_\infty+ \left(\frac{J}{2}+\frac{b^2a_\infty}{2}\log(\sqrt{B} u)\right)u^2 + \cdots\ .
    \end{split}
    \label{theta_a_expand}
\end{equation}
We have chosen the constants in the logarithms such that their arguments become scale-invariant.
This can always be done by a redefinition of $C$ and $J$. Expanding the DBI action at infinity ($u\sim0$), we obtain
\begin{multline}
\frac{S}{\mathcal{N} V_4} = \int^\epsilon du\left(\frac{1}{u^5}-\frac{M^2}{u^3}+\frac{a_\infty k^2}{2u}+\frac{b^2 M^2}{2u}+...\right)\\
=-\frac{1}{4\epsilon^4}+\frac{M^2}{2\epsilon^2}+\frac{a_\infty ^2k^2}{2}\log\epsilon+b^2M^2\log\epsilon+\cdots
\end{multline}
where we have introduced a UV cutoff at $u=\epsilon$. The counterterms required to regularize the on-shell action are given in \cite{Karch:2005ms,Karch:2006bv,Hoyos:2011us,Fadafan_2021},
\begin{multline}
    \frac{S_{ct}}{\mathcal{N}}=\int d^4x \sqrt{-\gamma}\bigg(
    \frac{1}{4}
    -\frac{1}{2}|\Theta|^2
    +\frac{5}{12}|\Theta|^4
    +\frac{1}{2}\sqrt{-\gamma}\log|\Theta|\Theta^*\square_{\gamma}\Theta\\
    +\frac{1}{4}\sqrt{-\gamma}\Theta^*\square_{\gamma}\Theta
    -\frac{1}{4}(2\pi\alpha')^2 \gamma^{\mu\rho} \gamma^{\nu\sigma} F_{\mu\nu}F_{\rho\sigma}\sqrt{-\gamma}\log (\epsilon/u_0)
    \bigg)~,
\end{multline}
with $\Theta=\theta e^{i\phi}$ and $\square_{\gamma}\Theta=\frac{1}{\sqrt{-\gamma}}\partial_{\mu}\left(\sqrt{-\gamma}\gamma^{\mu\nu}\partial_{\nu}\Theta\right)$.
A constant $u_0$ is introduced to make the argument of the logarithm scale-invariant. The total contribution from the counterterms is
\begin{equation}
    \frac{S_{ct}}{\mathcal{N}V_4}=\frac{1}{4\epsilon^4}-\frac{M^2}{2\epsilon^2}-b^2M^2\log(M\epsilon)-MC+\frac{M^2}{4}(M^2-b^2)-\frac{k^2a_\infty^2}{2}\log (\epsilon/u_0)\ ,
    \label{Sct}
\end{equation}
correctly canceling the divergences when $\epsilon\to 0$. 

\subsection{Quark condensate and electric current}

We now consider the infinitesimal variations of the parameters: $M\to M+\delta M$ and $a_\infty\to a_\infty+\delta a_\infty$.  As a result, the solution of the brane is also deformed as $\theta\to\theta + \delta\theta$ and $a\to a+\delta a$. The variation of the DBI action becomes a boundary term, given by
\begin{equation}
    \frac{\delta S}{\mathcal{N}V_4}=\delta \theta \frac{\partial \mathcal{L}}{\partial \theta'}\bigg|_{u=0} + \delta a \frac{\partial \mathcal{L}}{\partial a'} \bigg|_{u=0}\ ,
\end{equation}
where we used the equations of motion for $\theta$ and $a$. Expanding the variations near infinity, using Eq. (\ref{theta_a_expand}), we obtain
\begin{equation}
\begin{split}
    &\delta\theta = \delta Mu + \left(\delta C+\frac{M^2}{2}\delta M+\frac{b^2\delta M}{2}\log(Mu)+\frac{b^2\delta M}{2}\right)u^3 + \cdots\ ,\\
    &\delta a = \delta a_\infty+ \left(\frac{\delta J}{2}+\frac{k^2\delta a_\infty}{2}\log(\sqrt{ka_\infty} u)+ \frac{k^2\delta a_\infty}{4}\right)u^2 + \cdots\ .
\end{split}
\end{equation}
Also using Eqs.~(\ref{SDBI_u}) and (\ref{theta_a_expand}), we have
\begin{equation}
\begin{split}
    &\frac{\partial \mathcal{L}}{\partial \theta'} = \frac{M}{u^3} + \left(\frac{b^2M}{2}-\frac{3m^3}{2}+3C+\frac{3b^2M}{2}\log(mu)\right)\frac{1}{u}  + \cdots\ ,\\
    &\frac{\partial \mathcal{L}}{\partial a'} = J + \frac{k^2a_\infty}{2}+k^2a_\infty \log(\sqrt{ka_\infty}u)\delta a_\infty + \cdots\ .
\end{split}
\end{equation}
Therfore, the variation of the DBI action is given by
\begin{multline}
    \frac{\delta S}{\mathcal{N}V_4}=\frac{M\delta M}{\epsilon^2} + \left(-M^3+b^2M+3C+2b^2M\log(M\epsilon)\right)\delta M\\
    + M\delta C
    +\left(J+\frac{k^2a_\infty}{2}+k^2a_\infty \log(\sqrt{ka_\infty} \epsilon)\right)\delta a_\infty + \cdots\ .
\end{multline}
On the other hand, the variation of $S_{ct}$ is 
\begin{multline}
    \frac{\delta S_{ct}}{\mathcal{N}V_4}=-\frac{M\delta M}{\epsilon^2}-2b^2M\delta M \log(M\epsilon)\\
    -\delta M C - M\delta C +M^3\delta M -\frac{3b^2M\delta M}{2}\\
    -k^2a_\infty\delta a_\infty \log (\sqrt{ka_\infty}\epsilon/\alpha)-\frac{k^2a_\infty}{4}\delta a_\infty\ ,
    \label{varSct}
\end{multline}
where we set $u_0=\gamma/\sqrt{ka_\infty}$, with $\gamma$ being an arbitrary constant. Finally, the variation of the on-shell action is
\begin{equation}
    \frac{\delta (S+S_{ct})}{\mathcal{N}V_4}=
    \left(2C-\frac{b^2M}{2}\right)\delta M + \left(J+\left(\frac{1}{4}+\log\gamma \right)k^2a_\infty\right)\delta a_\infty\ .
\end{equation}
Thus, the quark condensate and electric current are obtained as 
\begin{equation}
\begin{split}
    &\langle \mathcal{O}\rangle_m|_\textrm{QFT} \equiv -\frac{2\pi\alpha'}{V_4}\frac{\delta S}{\delta M}=\frac{N_cN_f\sqrt{\lambda}}{8\pi^3}\left(-2C+\frac{b^2M}{2}\right)\ ,\\
    &J_x+iJ_y |_\textrm{QFT} \equiv  {2\pi \alpha'} \frac{\delta S}{\delta a_\infty}e^{ikz}=\frac{N_cN_f\sqrt{\lambda}}{8\pi^3}\left(J+\left(\frac{1}{4}+\log\gamma \right)k^2a_\infty\right)e^{ikz}\ .
    \end{split}
\end{equation}
Here, we use $\alpha'=1/\sqrt{\lambda}$ and $\mathcal{N}=N_cN_f \lambda/(2\pi)^4$. The term $\log\gamma$ represents an ambiguity in the current, arising from the ambiguity in the local counterterm. In the main text, we set $\gamma=1$. 

\subsection{Technical detail for computing free energy}

We introduce a UV-cutoff at $r=r_c$ in the original coordinate used in Eq.~(\ref{AdS5S5}). The relation between $\epsilon$ and $r_c$ is given by
\begin{equation}
    \frac{1}{\epsilon}
    =r_c+\frac{M^2}{2r_c^2}
+\frac{M}{r_c^3}(C-\frac{M^3}{8})
-\frac{b^2M^2}{2r_c^3}\log(r_c/M)+\cdots\ .
\end{equation}
Substituting this expression into Eq.(\ref{Sct}), we obtain the action of the counterterms,
\begin{equation}
    \frac{S_{ct}}{\mathcal{N} V_4}=\frac{r_c^4}{4}-\frac{b^2M^2}{4}+\frac{b^2M^2}{2}\log(r_c/M) 
    +\frac{k^2a_\infty^2}{2}\log(u_0r_c)\ .
    \label{Sct_rc}
\end{equation}
In principle, we can compute the free energy~(\ref{fdef}) using the above expression as the counterterm. However, in actual numerical calculations, there is a cancellation of large values and the truncation error is not negligible. To reduce the truncation error, we include the counterterm~(\ref{Sct_rc}) in the integrand. We define 
\begin{equation}
    \mathcal{L}_{ct}(r)=\frac{r^4}{4}-\frac{b^2R^2(r)}{4}+\frac{b^2R^2(r)}{2}\log(r) 
    +\frac{k^2a^2(r)}{2}\log(r)\ .
\end{equation}
Then, the regularized on-shell action written as
\begin{equation}
    \frac{S_{DBI}+S_{ct}}{\mathcal{N}V_4}=\int_0^{r_c} dr [\mathcal{L}_0+\{h(r)\mathcal{L}_{ct}(r)\}']-\frac{b^2M^2}{2}\log M  +\frac{k^2a_\infty^2}{2}\log u_0\ ,
\end{equation}
where the function $h(r)$ approaches $0$ near the origin and $1$ at infinity quickly enough. For example, we can choose $h(r)=r^6/(1+r^6)$.



\section{Discrete scale invariance}\label{app:discretescale}
In this appendix we study in more detail the spiral structure of the condensate observed in Fig.~\ref{fig:c_vs_m}. Let $R_{0n}$ ($n = 0,1,2,\cdots$) represent the values of $R_0$ where the quark mass $M$ vanishes, arranged in descending order. The brane profile $R = R(r)$ corresponding to $R_0 = R_{0n}$ intersects the horizontal axis $n$ times (see Fig. \ref{fig:c_vs_m}). We will refer to such a solution as the $n$-node solution.

Now, let's examine the self-similarity structure near the origin in more detail. As $R_{0}\to 0$, $R(r)$ becomes sufficiently small, while $a(r)$ remains finite. We linearize the equations by substituting $R(r) \to \epsilon R(r)$ and considering the first-order term in $\epsilon$. This leads to the linearized equation for $R(r)$ near $r=0$,
\begin{equation}
    R'' + \frac{R'}{r} + \frac{2R}{r^{2}}=0.
    \label{eq:smallR}
\end{equation}
One can easily check that the equation has the scaling symmetry $R(r)\to R(\mu r)/\mu$, with $\mu$ a real positive constant. The solution of (\ref{eq:smallR}) is given by $R \sim r^{\nu_{\pm}}$, with a purely imaginary exponent $\nu_{\pm }=\pm\sqrt{2} i$, or equivalently,
\begin{equation}
    R(r) = C_{1} \sin \left( \sqrt{2}\log r \right) + C_{2} \cos\left( \sqrt{2}\log r \right),
\end{equation}
where $C_1$ and $C_{2}$ are constants. The scaling symmetry of the equation leads to the coefficients transformation
\begin{equation}
    \begin{pmatrix}
C_{1} \\
C_{2} 
\end{pmatrix}
\to
\frac{1}{\mu}
\begin{pmatrix}
\cos(\sqrt{2} \log \mu) & -\sin(\sqrt{2} \log \mu) \\
\sin(\sqrt{2} \log \mu) & \cos(\sqrt{2} \log \mu)
\end{pmatrix}
\begin{pmatrix}
C_{1} \\
C_{2} 
\end{pmatrix}.
\end{equation}
This scaling symmetry and transformation imply that the brane embedding shows discrete self-similarity with a period $\mu_{2\pi} = \exp (\sqrt{2}\pi)$.
Although this result comes from the analysis near $r=0$, it naturally explains the oscillatory behavior of the quantities at the boundary, such as the spiral behaviour in Fig.~\ref{fig:c_vs_m}. 

\begin{figure}[t]
\centering
\includegraphics[width=0.7\textwidth]{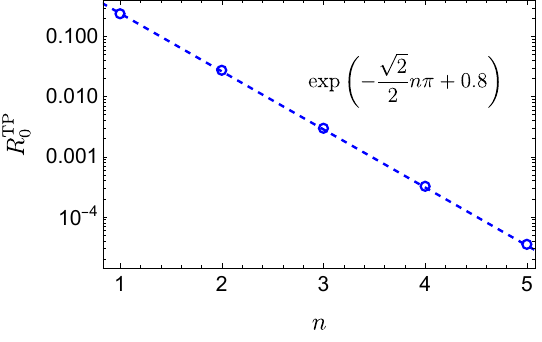}
\caption{The $n$-th turning point in the spiral behaviour of Fig.~\ref{fig:c_vs_m} and the corresponding value of $R_{0}\equiv R_{0}^{\rm TP}$. The dashed line represents the semi-analytical predictions for the period based on the discrete self-similarity.}
\label{fig:R0TP}
\end{figure}

To illustrate this, we plot the $n$-th turning point in the spiral behaviour of Fig.~\ref{fig:c_vs_m}, and the corresponding value of $R_{0}\equiv R_{0}^{\rm TP}$ in Fig.~\ref{fig:R0TP}.

Since $R_{0}$ scales as $R_{0} \to R_{0}/\mu$, each turning point is expected to appear at a half-period:~$\mu_{\pi}=\exp(\sqrt{2}\pi/2)$. Hence, we obtain $R_{0}^{\rm TP} \propto \exp(-\sqrt{2}n\pi/2)$. As shown in the dashed line in Fig.\ref{fig:R0TP}, we confirm that the turning point of the spiral in the quark mass and condensate is well described by the self-similarity of the brane embedding.

A similar analysis has been performed near the critical embedding at finite temperature \cite{Frolov:2006tc,Mateos:2006nu,Mateos:2007vn}, or in the presence of a finite electric field \cite{Ishigaki:2023dss}. However, in our case, the self-similarity appears near the trivial solution $R=0$, rather than near the critical embedding with the conical singularity at the black hole horizon or effective horizon created by an external electric field.

\bibliography{refs}

\end{document}